\documentclass{iopart}              

\usepackage{iopams}
\usepackage{graphicx}
\usepackage{cite}

\begin{document}

\title[]{A New Multiferroic Material: MnWO$_4$}

\author{O Heyer$^1$, N Hollmann$^1$, I Klassen$^1$, S Jodlauk$^2$, L Bohat\'{y}$^2$, P Becker$^2$, J A Mydosh$^1$, T~Lorenz$^1$ and
D~Khomskii$^{1,3}$\footnote[0]{$^3$ Author to whom correspondence
should be addressed (khomskii@ph2.uni-koeln.de).} }

\address{$^1$ II.\,Physikalisches Institut, Universit\"{a}t zu K\"{o}ln, Z\"{u}lpicher Str.
77, 50937 K\"{o}ln, Germany}

\address{$^2$ Institut f\"{u}r Kristallographie,
Universit\"{a}t zu K\"{o}ln, Z\"{u}lpicher Str. 49b, 50674
K\"{o}ln, Germany}

\begin{abstract}
We report the multiferroic behaviour of MnWO$_4$, a magnetic oxide
with monoclinic crystal structure and spiral long-range magnetic
order. Based upon recent theoretical predictions MnWO$_4$ should
exhibit ferroelectric polarization coexisting with the proper
magnetic structure. We have confirmed the multiferroic state below
13~K by observing a finite electrical polarization in the
magnetically ordered state via pyroelectric current measurements.
\end{abstract}





Multiferroic materials which combine magnetism and
ferroelectricity currently attract considerable
attention~\cite{fiebig04b,fiebig05a,khomskii06a,tokura06a}. There
are already several multiferroic materials recently discovered
among transition metal oxides: TbMnO$_3$~\cite{kimura03a},
TbMn$_2$O$_5$~\cite{hur04a}, DyMnO$_3$~\cite{kimura05a}.
Nevertheless, the search for novel systems with multiferroic
properties presents a definite interest. In this letter we report
that yet another transition metal oxide, MnWO$_4$, belongs to the
same class of materials and develops spontaneous electric
polarization in a spiral magnetically ordered
state~\cite{Kimura06too}.

There exist several different microscopic mechanisms which may
cause multiferroic behavior~\cite{khomskii06a}. One of the most
interesting cases is when a spontaneous polarization exists only
in a magnetically ordered phase with a particular type of
ordering. This is e.g.\ the case in TbMnO$_3$ and TbMn$_2$O$_5$.
Microscopic~\cite{katsura05a} and
phenomenological~\cite{mostovoy06a} treatments have shown that
this happens particularly in spiral magnetic structures with the
spin rotation axis $\overrightarrow{e}$ not coinciding with the
magnetic propagation vector $\overrightarrow{Q}$: theoretical
treatment shows that in this case a finite spontaneous
polarization perpendicular to the plane spanned by
$\overrightarrow{e}$ and $\overrightarrow{Q}$ may appear
\begin{equation}
\overrightarrow{P}\sim \overrightarrow{e} \times
\overrightarrow{Q} \,\, . \label{PM}
\end{equation}
This is not the only source for a magnetically driven
ferroelectricity~\cite{aliouane06a,sergienko06b}, but perhaps the
most common one. Accordingly, one strategy to search for new
multiferroic materials is to look for magnetic systems with proper
magnetic structures. MnWO$_4$ (also known as the mineral
h\"{u}bnerite) appears to be just such a system. Detailed studies
of the magnetic ordering in this material have
shown~\cite{lautenschlager93a, ehrenberg97a} that below 12.3~K a
spiral magnetic ordering develops which seems to satisfy the
criterion of Eq.~(\ref{PM}). In order to test this we carried out
measurements of the dielectric response and of spontaneous
polarization of MnWO$_4$ using single-crystalline samples.

\begin{figure}[t]
\begin{center}
\includegraphics[width=0.8\textwidth]{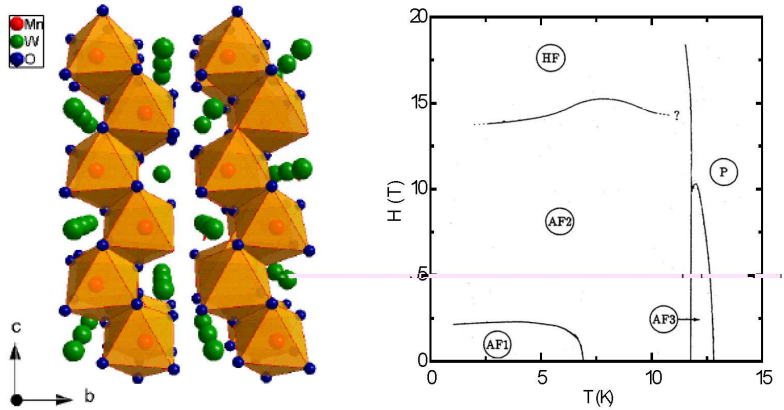}
\end{center}
\caption{\label{struc} Crystal structure of MnWO$_4$ and
schematic $H-T$ phase diagram for a magnetic field applied along
the easy axis according to Refs.~\cite{lautenschlager93a}
and~\cite{ehrenberg97a}, respectively.}
\end{figure}

The crystals of MnWO4 were grown from melt solution. On the basis
of earlier work~\cite{schultze67a} we applied a modified flux
technique, using a melt solvent from the system Na$_2$WO$_4$ -
WO$_3$. The resulting crystals are of dimensions up to 15 x 5 x
3~mm$^3$ and of dark brown color. The crystal structure of
MnWO$_4$ is monoclinic (space group of the paramagnetic phase
P2/c) and consists of edge-sharing [MnO$_6$] and [WO$_6$]
octahedra that form zig-zag chains along the c-axis, see
Fig.~\ref{struc}. Tungsten atoms and manganese atoms are arranged
in alternating sheets parallel to (100)~\cite{weitzel76a}.

There is apparently also a finite inter-chain coupling causing the
observed magnetic ordering below $\simeq 13$~K (see
Fig.~\ref{struc})~\cite{lautenschlager93a,ehrenberg97a}. According
to the previous results there are two separate transitions at
13.5~K and 12.3~K. The upper one is from the paramagnetic state to
an incommensurate sinusoidal spin-density wave state, with wave
vector $\overrightarrow{Q}=(-0.214,1/2,0.457)$. The spins are
collinear in the $ac$ plane with an angle of about $35^\circ$ with
respect to the $a$ axis. For later use, this direction is termed
easy axis. In the so-called AF2 phase below 12.3~K the wave vector
hardly changes, but a finite spin component moment along the $b$
axis develops and, as a consequence, an elliptical spiral
structure evolves. Finally, a transition to a commensurate
magnetic structure with $Q=(\pm 1/4,1/2,1/2)$ is found around 8~K,
in which the spins are again collinear. Our magnetic measurements,
see below, confirm the presence of at least two magnetic
transitions, one around 13~K and a second one around 6~K, but we
were unable to resolve two separate transitions at 12.3 and
13.5~K. According to Refs.~\cite{lautenschlager93a,ehrenberg97a}
the separate transitions around 13~K can probably be much better
resolved by e.g.\ specific heat measurements than in magnetic
data. Moreover, it is reported that the transition to the AF1
phase is of first order and is found between 6.8 and 8~K for
different samples and measurement techniques.

With respect to multiferroic behavior, the AF2 phase is the one of
interest, because the magnetic structure is a spiral with spin
rotation axis $\overrightarrow{e}$ different from the wave vector
$\overrightarrow{Q}$, which according to Eq.~(\ref{PM}) should
lead to ferroelectricity. The direction of $\overrightarrow{e}$ is
given by the cross product of the above-mentioned easy axis and
the $b$ axis, and via Eq.~(\ref{PM}) we expect a finite
polarization in the plane spanned by the easy axis and the b axis
with an angle of $\simeq$10$^\circ$ with respect to the easy axis.

We used a sample with rectangular (100) surfaces of about $6
\times 4$~mm$^2$ and a thickness of about 0.7~mm. We have chosen
this orientation, since the morphology of our MnWO$_4$ crystals is
dominated by (100). Gold electrodes have been sputtered onto the
opposite faces of the sample, and the dielectric constant
$\varepsilon$ has been determined by measuring the capacitance of
the sample using a precision capacitance bridge (Andeen-Hagerling
2500A). To obtain the polarization $\overrightarrow{P}$, we
measured  the pyroelectric current $I_p$ using an electrometer
(Keithley 6517A), while sweeping the temperature of the sample at
a rate of $\sim 2$~K/min. To avoid domain formation with opposite
directions of $\overrightarrow{P}$, we applied an electric field
(300~V/mm) while cooling the sample from a temperature well above
$T_{N}\simeq 13$~K. The electric field was removed before the
$I_p(T)$ measurements during the heating process. The polarization
was determined by integrating the pyroelectric current with
respect to the time.

\begin{figure}[t]
\begin{center}
\includegraphics[width=1\textwidth]{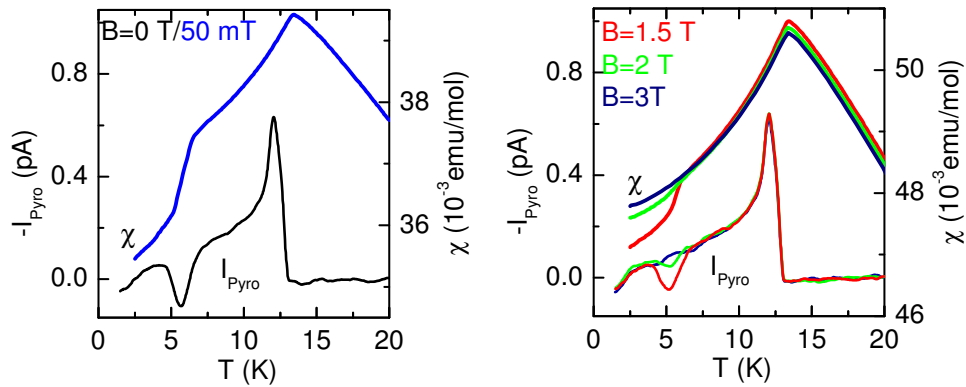}
\end{center}
\caption{\label{pyroc} Pyroelectric current $I_p$ (measured along
[100], left axis) and magnetic susceptibility $\chi$ (right axis)
of MnWO$_4$ measured in zero (or very low) field (left panel) and
in higher magnetic fields applied  along the c axis.}
\end{figure}

In our measurements of $\varepsilon(T)$ we observed various
anomalies in the vicinity of $\simeq 12$~K, i.e.\ close to the
magnetic ordering, but the overall change of $\varepsilon(T)$
remained rather small. The results of the pyroelectric current
measurements shown in Fig.~\ref{pyroc} reveal that the magnetic
transition around 13~K is accompanied by a peak in the
pyroelectric current. With further decreasing temperature, this
current continuously decreases until another peak of opposite sign
occurs around 6~K. The comparison of $I_p(T)$ in zero magnetic
field with the magnetic susceptibility $\chi(T)$ measured in a
field of 50~mT shows that the anomalies in $I_p(T)$ are clearly
related to the magnetic transitions from the PM to the AF3/2 phase
and from the AF2 to the AF1 phase, respectively. This correlation
is further confirmed by measurements in magnetic fields up to 3~T.
As shown in Fig.~\ref{pyroc}, both $\chi(T,H)$ and $I_p(T,H)$ do
hardly change with field in the temperature range above 7~K, while
the additional anomalies around 6~K simultaneously decrease with
increasing field and vanish for both quantities in a field of
about 3~T. In our measurements the magnetic field has been applied
along the c axis which means that the contribution of the applied
field parallel to the easy axis is $\sim$55\%. According to
Ref.~\cite{ehrenberg97a} the transition from the AF2 to the AF1
phase is suppressed in fields above about 2~T for this field
direction (see Fig.~\ref{struc}). This naturally explains the
vanishing of the low-temperature anomalies in both, $\chi(T,H)$
and $I_p(T,H)$ for $H\geq 3$~T.

\begin{figure}[t]
\begin{center}
\includegraphics[width=0.8\textwidth]{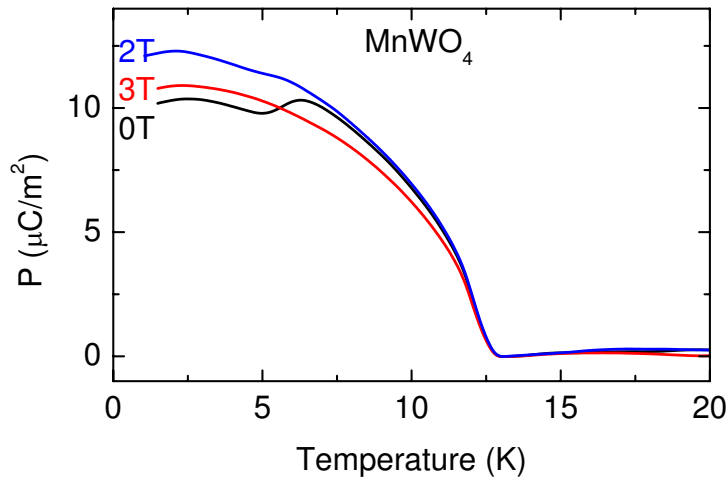}
\end{center}
\caption{\label{pol} Polarization along the $a$ axis of MnWO$_4$
for different magnetic fields applied along the c axis.}
\end{figure}

In Fig.~\ref{pol} we show the spontaneous polarization calculated
via $P(T)=\int I_p(T,t) dt/A$ ($A=24$mm$^2$ is the sample
surface). Obviously, $P(T)$ continuously increases with decreasing
$T\lesssim 12.5$~K as it is expected for a standard second-order
phase transition. The maximum value of $P$ is of the order of only
10$\mu$C/m$^2$. In our geometry, we measure the projection of
$\overrightarrow{P}$ onto the $a$ axis. The angle between $a$ and
the direction of $\overrightarrow{P}$ expected from Eq.~(\ref{PM})
is about 36$^\circ$. Thus, one may expect that the actual value of
$\overrightarrow{P}$ is about 15$\mu$C/m$^2$, which is more than
one order of magnitude smaller than  the values observed in other
multiferroic materials, as e.g.\ TbMnO$_3$~\cite{kimura03a}, and
about 4 orders of magnitude smaller than $\overrightarrow{P}$ of
typical ferroelectric, e.g.\ BaTiO$_3$~\cite{jona62a}.

According to Ref.~\cite{lautenschlager93a}, the magnetic structure
in the AF1 phase is collinear. From Eq.~(\ref{PM}), one should
therefore expect a vanishing $P$ in the AF1 phase, while we
observe only a partial decrease of $P$ at the AF2-to-AF1
transition in our experiment. One possible explanation for this
could be that the finite $P$ arises from one of the other
mechanisms, which have been proposed to explain
ferroelectricity~\cite{aliouane06a, sergienko06b} and may also
work within a collinear phase. Alternatively, we cannot exclude
that there may be a small content of impurities (e.g. Mn$_3$O$_4$)
in our sample, which causes only a partial transformation from the
spiral and ferroelectric phase AF2 to the collinear phase AF1. Due
to the first-order nature of the AF1-AF2 transition, we consider
such a phase coexistence as rather plausible, but in order to
clarify the real nature of the low-temperature phase further
experiments are needed. Moreover, I$_P$ measurements along
different axes have to be performed in order to determine
experimentally the direction of P.

In summary, we have shown that there appears a spontaneous
electric polarization in MnWO$_4$ below $\sim 13$~K in the
magnetically ordered phase, whose magnetic structure is described
by an elliptical spiral. Thus, MnWO$_4$ is yet another
multiferroic transition metal oxide.

\ack

We are very grateful to M.~Mostovoy for useful discussions. This
work was supported by the DFG through SFB 608. The work of D.~Kh.\
was also supported by the MRSEC Grant DMR-0520471.

\section*{References}


\end{document}